\documentclass[aps,prb,twocolumn,showpacs,superscriptaddress]{revtex4}

\usepackage{graphicx}
\usepackage{amsmath}
\usepackage{soul}

\usepackage[colorlinks,linkcolor=blue,anchorcolor=blue,citecolor=blue]{hyperref}

\begin{document}

\title{Neutron scattering studies on ionic diffusion behaviors of superionic $\alpha$-Cu$_{2-\delta}$Se}
\author{Lisi Li}
\affiliation{Center for Neutron Science and Technology, Guangdong Provincial Key Laboratory of Magnetoelectric Physics and Devices, School of Physics, Sun Yat-Sen University, Guangzhou, 510275, China}
\author{Huili Liu}
\affiliation{School of Physical Science and Technology, ShanghaiTech University, Shanghai, 201210, China}
\author{Maxim Avdeev}
\affiliation{Australian Nuclear  Science and Technology Organisation,New Illawarra Road, Lucas Heights NSW 2234, Australia.}
\author{Dehong Yu}
\affiliation{Australian Nuclear  Science and Technology Organisation,New Illawarra Road, Lucas Heights NSW 2234, Australia.}
\author{Sergey Danilkin}
\affiliation{Australian Nuclear Science and Technology Organisation,New Illawarra Road, Lucas Heights NSW 2234, Australia.}
\author{Meng Wang}
\email{wangmeng5@mail.sysu.edu.cn}
\affiliation{Center for Neutron Science and Technology, Guangdong Provincial Key Laboratory of Magnetoelectric Physics and Devices, School of Physics, Sun Yat-Sen University, Guangzhou, 510275, China}

\begin{abstract}
We present studies on crystal structure and ionic diffusion behaviors of superionic Cu$_{2-\delta}$Se ($\delta$=0, 0.04, and 0.2) by utilizing neutron powder diffraction and quasi-elastic neutron scattering. In the superionic phase, the structural model with Cu ions occupying the Wyckoff sites of 8$c$ and 32$f$ provides the best description of the structure. As the content of Cu increasing in Cu$_{2-\delta}$Se, the Cu occupancy increases on the 32$f$ site, but decreases on the 8$c$ site . Fitting to the quasi-elastic neutron scattering spectra reveals two diffusion modes, the localized diffusion between the 8$c$ and 32$f$ sites and the long-range diffusion between the adjacent 8$c$ sites using the 32$f$ site as a bypass, respectively. Between 430 and 650 K, we measured that the compound with more Cu content exhibits a larger long-range diffusion coefficient. Temperature in this range does not affect the long-range diffusion process obviously. Our results suggest the two diffusion modes cooperative and thus provide a microscopic understanding of the ionic diffusion of the Cu ions in superionic Cu$_{2-\delta}$Se.
\end{abstract}

\maketitle

\begin{figure*}[t]
\centering
\includegraphics[width=16 cm]{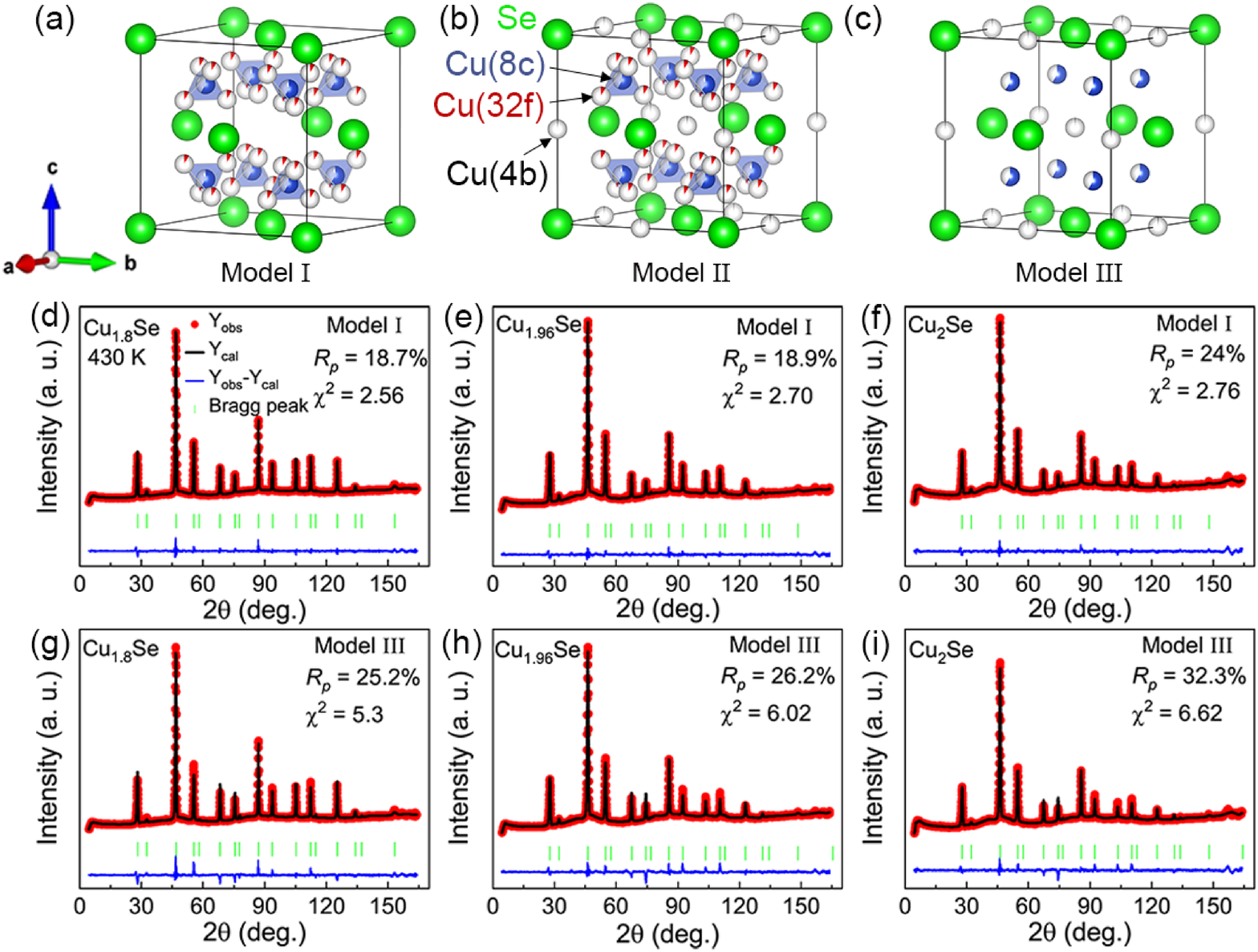}\\
\caption{(a) Structural models for describing $\alpha$-Cu$_{2-\delta}$Se. Model I: Cu ions occupy the $8c$ and $32f$ sites. (b) Model II: Cu ions occupy the $8c$, $32f$, and $4b$ sites. (c) Model III: Cu ions occupy the 8c and 4b sites. (d) Observed and calculated neutron diffraction patterns at 430 K for Cu$_{1.8}$Se, (e) Cu$_{1.96}$Se, and (f) Cu$_2$Se using Model I. (g-i) Comparisons between the identical observed neutron diffraction patterns and calculated patterns using the Model III. The refinement agreement factors $R_p$ and $\chi^2$ have been listed on the figures (d-i).}
\label{fig1}
\end{figure*}

Thermoelectric (TE) materials that could convert heat to electricity provide a way for waste heat recycle\cite{He2017}. The dimensionless figure of merit $\emph{zT}=S^2\sigma T/\kappa$ is a measure of the efficiency of TE materials in application, where $S$, $\sigma$, $T$, and $\kappa$ are the Seebeck coefficient, electrical conductivity, temperature, and thermal conductivity, respectively\cite{Ioffe1959,Tritt2006}. The $\emph{zT}$ values for high performance TE materials can be higher than 2\cite{He2017}, for instance, SnSe\cite{Zhao2014}, PbTe\cite{Hsu2004,Heremans2008}, CoSb$_3$\cite{Rogl2014}, and Cu$_2$Se\cite{Liu2012,Li2018}. To obtain an ultra-low thermal conductivity is an important way to achieve a high $\emph{zT}$ value for TE materials\cite{Liu2012,Zhao2014}.

Cu$_{2-\delta}$Se is a high performance TE material, hosting a high temperature superionic $\alpha$-phase and a low temperature localized $\beta$-phase with a phase transition at 414 K for $\delta=0$\cite{Tonejc1980,Korzhuev1989a,Danilkin2009,Zhao2020}. The transition temperature decreases as the increasing of the Cu deficiency $\delta$. In the stabilized $\beta$-phase, Cu$_{2-\delta}$Se crystallizes into a monoclinic structure, where the Cu ions form a localized order structure. In the superionic $\alpha$-phase, Cu$_{2-\delta}$Se forms a cubic structure with the space group $Fm\bar{3}m$ (No. 225). The $\emph{zT}$ value can reach up to 2.5 for Cu$_2$Se due to the ultra-low thermal conductivity\cite{Liu2012,Li2018}. The ultra-low thermal conductivity may relate to the ionic diffusion behaviors of the Cu ions in Cu$_{2-\delta}$Se\cite{Liu2012}. It is therefore important to investigate the diffusion behaviors of Cu$_{2-\delta}$Se for elucidation the relation between the Cu diffusion and the ultra-low thermal conductivity. However, there are significant discrepancies among the experimental determined diffusion coefficients ($D$), which are an important parameter for evaluating the diffuse behaviors. The diffusion coefficient was determined to be on the scale of 10$^{-7}$ cm$^2$/s using quasi-elastic neutron scattering (QENS)\cite{Voneshen2017}, smaller by two orders of magnitude than the values determined by chemical and other QENS experiments\cite{Yakshibaev1984,Korzhuev1989,Danilkin2011,Danilkin2012,NazrulIslam2021,Kumar2022}. The discrepancies in diffusion coefficient may originate from different diffusion models in analyzing the QENS data. In addition, the specific diffusion paths of the Cu ions are under debates. Therefore, careful analyses of the diffusion coefficient is crucial for understanding the diffusion process and the mechanism of the ultra-low thermal conductivity in superionic Cu$_{2-\delta}$Se.

Here we present detailed investigations on the crystal structure and diffusion behaviors of superionic Cu$_{2-\delta}$Se ($\delta=0, 0.04, 0.2$) using neutron powder diffraction (NPD) and QENS. The refined averaged structure can be described well by a structural model with two Wyckoff sites of 8$c$ (1/4, 1/4, 1/4) and 32$f$ $(x, x, x)$ ($0.331\le x\le 0.337$) for Cu ions [Fig. \ref{fig1}]. The QENS signals arising from the diffusion of Cu ions are observed in the superionic phase. Through our QENS data, a long-range jump diffusion mode and a localized confined mode are obtained. For the long-range diffusion mode, the residence time $\tau$ and averaged jump length $l$ are extracted from the Chudley-Elliott (C-E) model\cite{Chudley1961}. The residence time $\tau$ is on the time scale of $10^{-12}$ s and becomes shorter for the higher Cu content. The averaged jump length $l$ is in the range of $2.7-3.4$ {\AA} that could capture the main jump paths of Cu ions diffusing between the adjacent $8c$ sites. In addition, the diffusion coefficient $D$ is on the order of magnitude $10^{-5}$ cm$^2$/s. The localized confined mode corresponds to the diffusion between the nearest 8$c$ and 32$f$ sites which is cooperative with the long-range diffusion mode. Our results reveal the influence of Cu content on the Cu ions diffusion process and demonstrate that the liquid-like mobilities of Cu ions are prominent in Cu$_{2-\delta}$Se.

Powder samples of Cu$_{2-\delta}$Se ($\delta=0, 0.04, 0.2$) were synthesized using the standard solid-state reaction method as in previous reports\cite{Danilkin2011,Danilkin2012,Liu2016}. Starting materials of Cu and Se powder were mixed in the stoichiometric quantities and sealed in quartz tubes under vacuum, which were then placed into a box furnace and heated at 1000 $^{\circ}$C for 7 days. The NPD data were collected on the high resolution powder diffractometer Echidna installed at the OPAL reactor of the Australian Nuclear Science and Technology Organisation (ANSTO)\cite{Avdeev2018}. The incident neutron wavelength was $\lambda$=1.622 {\AA}. The Rietveld method was employed to analyze the NPD data using the $FullProf$ Suite software\cite{Rietveld1969}.

The QENS experiments were conducted on the time-of-flight spectrometer Pelican at ANSTO with an incident neutron wavelength of $\lambda$=4.69 {\AA} (3.72 meV) and different chopper configurations\cite{Yu2013}. The powder samples were loaded into sealed aluminum sample cans. The measured temperatures were 430, 550, and 650 K for Cu$_{1.8}$Se and Cu$_{1.96}$Se, and 430 and 650 K for Cu$_{2}$Se. An empty aluminum sample can was measured at the same temperatures for background subtraction. The spectra of a vanadium cylinder with 1 mm thick were measured to serve as resolution functions in analyzing the QENS spectra using the program $LAMP$\cite{Richard1996}. The energy resolutions determined by the full width at half maximum (FWHM) of the elastic peak of vanadium at $Q$ = 1.05 $\pm$ 0.15 {\AA} is 207.8(9) $\mu$eV for the earlier measurements on Cu$_{2}$Se and 138.9(5) $\mu$eV for the later measurements on Cu$_{1.8}$Se and Cu$_{1.96}$Se.

\begin{figure*}
\centering
\includegraphics[width=16cm]{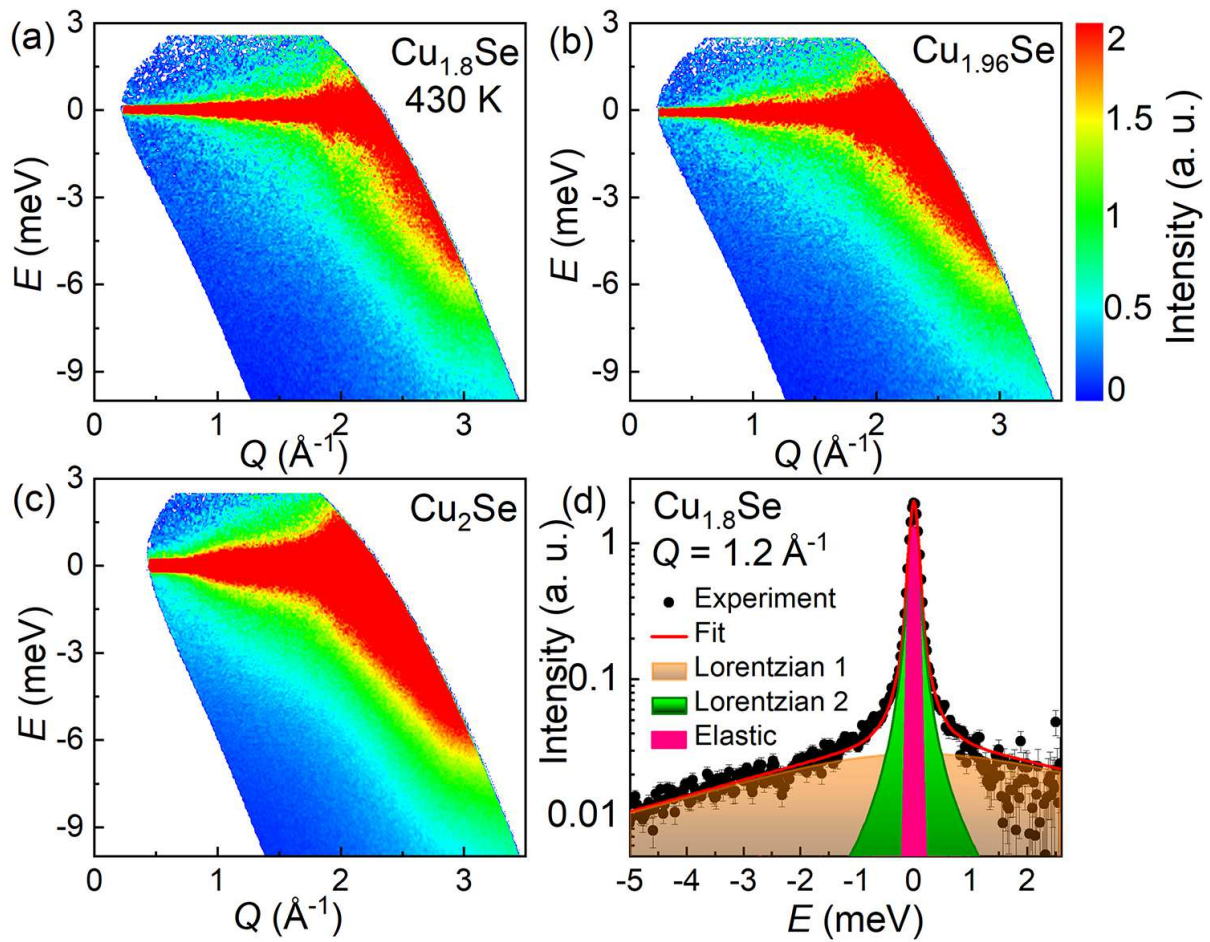}\\
\caption{The QENS spectra measured at 430 K for (a) Cu$_{1.8}$Se, (b) Cu$_{1.96}$Se, and (c) Cu$_2$Se. The color represents intensities. (d) Black symbols are a constant $Q$ cut at $Q=1.2$ {\AA}$^{-1}$ integrated within $Q\pm0.075$ {\AA}$^{-1}$ at 430 K for Cu$_{1.8}$Se. The red solid line represents a fitting result of the QENS peak. The orange, green, and pink areas represent the Lorentzian 1, Lorentzian 2, and elastic components of the fitting.}
\label{fig2}
\end{figure*}

The Cu ions of Cu$_{2-\delta}$Se can diffuse through the Se cubic lattice in the superionic $\alpha$-phase\cite{MILAT1987,Kashida1988,Frangis1991}. Diffraction data based on structure refinement can only capture the time- and space-averaged positions of Cu ions in this phase. Three structural models have been proposed for describing the distribution of the Cu ions in the superionic phase, as shown in Figs. \ref{fig1}(a)-\ref{fig1}(c). Model I is that the Cu ions occupy the tetrahedral sites 8$c$ (1/4,1/4,1/4) and trigonal sites 32$f$ $(x,x,x)$ $(x=0.33-0.39)$\cite{Heyding1976,Skomorokhov2006,Danilkin2011,Danilkin2012}. Model II states that most of the Cu ions occupy the tetrahedral sites 8$c$ and the trigonal sites 32$f$, and a small amount of Cu ions sit in the octahedral sites 4$b$ (1/2,1/2,1/2)\cite{Boyce1981,Oliveria1988,Yamamoto1991,Yamamoto1991a}. Model III allows the Cu ions to occupy the sites of $8c$ and $4b$, while the $32f$ sites are not occupied\cite{NazrulIslam2021}.

\begin{table}
\caption{The nominal and refined compositions, the occupancies (Occ.) of Cu ions on the $8c$ and $32f$ sites, and the lattice constant $a$ using the Model I at 430 K for $\alpha$-Cu$_{2-\delta}$Se ($\delta=0, 0.04, 0.2)$.}
\setlength{\tabcolsep}{1.5mm}
\begin{tabular}{cccccc}
\hline \hline
Nominal &Refined &Occ.(8c) & Occ.(32f) &$\emph{a}$ ({\AA})  \\ \hline
Cu$_{1.8}$Se  &Cu$_{1.72(5)}$Se &0.62(1) &0.060(4) &5.7761(1)  \\
Cu$_{1.96}$Se &Cu$_{1.83(5)}$Se &0.58(1)&0.084(4)&5.8387(2)  \\
Cu$_{2}$Se &Cu$_{1.84(5)}$Se &0.59(1) &0.083(4)&5.8465(3)  \\ \hline
\end{tabular}
\label{table1}
\end{table}

To investigate the occupancy of the Cu ions, we conducted NPD measurements. Figures \ref{fig1}(d)-\ref{fig1}(f) show the NPD and refined patterns for Cu$_{2-\delta}$Se ($\delta=0, 0.04, 0.2$) at 430 K in the superionic $\alpha$-phase using the Model I. The refinement reveals a large amount of Cu ions occupying the 8$c$ sites and a small amount on the 32$f$ sites. The occupancy on the $8c$ sites is anticorrelated with the total amount of the Cu content, while the occupancy on the $32f$ sites increases for the compounds with higher Cu content. Table \ref{table1} summarizes the structure refinement results at 430 K for $\alpha$-Cu$_{2-\delta}$Se. The position parameters of Cu(32$f$) ions are $x=0.331(1)$, 0.332(1), and 0.337(1) for Cu$_{1.8}$Se, Cu$_{1.96}$Se, and Cu$_2$Se, respectively, in agreement with the reported results in the range of $x=0.33-0.39$\cite{Tonejc1980,Heyding1976,Skomorokhov2006,Danilkin2011,Danilkin2012}. The refined content of Cu ions as shown in Table \ref{table1} slightly deviates from the nominal quantity. Refinement results using the Model III are shown in Figs. \ref{fig1}(g)-\ref{fig1}(i). The refinement agreement parameters are worse than those of the Model I. In addition, the refinement will result in an unreasonable thermal vibration factor B $\sim40$ \AA$^2$ for the Cu ions on the $4b$ sites. A typical value of B should be below 5 \AA$^2$. The simulated NPD patterns using the Model II are very close to that of Model I. There is one additional freedom of the Cu sites for the Model II. The refinement agreement is expected to be better. However, the refinement using the Model II also yields an unreasonable B $\sim30$ \AA$^2$ for the Cu ions on the $4b$ sites. When the magnitude of B is enforced to be below 5 \AA$^2$, the resultant occupancy of Cu on the $4b$ sites is in the range of $1\sim3$\%. Thus, our refinements reveal that the Model I agrees with the structure of the high temperature superionic $\alpha$-phase Cu$_{2-\delta}$Se the best.

\begin{figure*}[t]
\centering
\includegraphics[width=17.5cm]{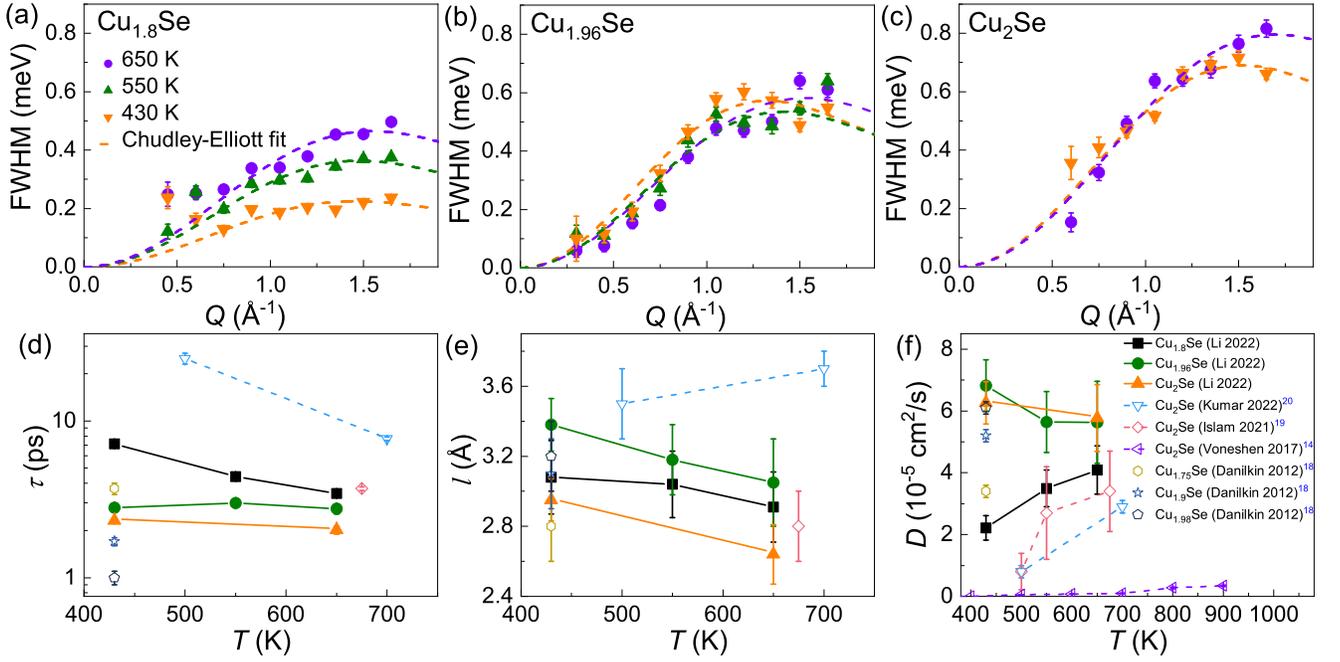}\\
\caption{$Q$ dependence of the FWHM for (a) Cu$_{1.8}$Se, (b) Cu$_{1.96}$Se, and (c) Cu$_2$Se measured at 430, 550, and 650 K. The dashed lines are fittings to the C-E function. Temperature dependence of (d) the residence time $\tau$, (e) the average jump length $l$, and (f) the diffusion coefficient $D$ in $\alpha$-Cu$_{2-\delta}$Se. The solid symbols refer to the results of this work. Results of (Kumar 2022) are reproduced with permission from Phys. Rev. Mater. 6, 055403 (2022)\cite{Kumar2022}. Copyright 2022 American Physical Society. Results of (Islam 2021) are reproduced with permission from Acta Mater. 215, 117026 (2021)\cite{NazrulIslam2021}. Copyright 2022 Elsevier. Results of (Voneshen 2017) are reproduced with permission from Phys. Rev. Lett. 118, 145901 (2017)\cite{Voneshen2017}. Copyright 2017 American Physical Society. Results of (Danilkin 2012) are reproduced with permission from Solid
State Ionics 225, 190 (2012)\cite{Danilkin2012}. Copyright 2012 Elsevier. }
\label{fig3}
\end{figure*}

We conducted QENS measurements on the same samples at 430, 550, and 650 K to investigate the diffusion behaviors microscopically. QENS experiments measure the dynamic process of individual atom such as rotation, reorientation, and diffusion in materials\cite{hempelmann2000,Embs2010}. The QENS spectra for $\alpha$-phase Cu$_{2-\delta}$Se ($\delta=0,0.04,0.2$) at 430 K are shown in Fig. \ref{fig2}. Strong broadening QENS signals around the elastic line are observed in all compounds, where Cu$_{2-\delta}$Se are in the superionic phase. The signals extend up to 9 meV in the energy gain side and the intense signals above $Q=2$ \AA$^{-1}$ should correspond to phonons of the samples.

The dynamic structure factor $S(Q, \omega)$ for the diffusion process has been fitted by the following formula\cite{hempelmann2000}:
\begin{equation}
S(Q,\omega)=[A_0(Q)\delta(\omega)+\sum _i A_i(Q)L_i(\omega)]\otimes R(Q,\omega),
\label{eq1}
\end{equation}
where $A_0(Q)$ is the incoherent elastic structure factor, $\delta(\omega)$ is the delta function, $A_i(Q)$ is a quasi-elastic component, and $L_i(\omega)$ is a Lorentzian function. The integer $i$ corresponds to different diffusion mode. The resolution function $R(Q,\omega)$ can be determined via the identical cuts from measurements on vanadium. The $S(Q, \omega)$ are fitted using two Lorentzian functions for all the momentum transfer $Q$ of the QENS data as in previous reports\cite{NazrulIslam2021,Kumar2022}. Figure \ref{fig2} (d) displays a constant $Q$ cut and the QENS peak fitting results at $Q=1.2\pm0.075${\AA}$^{-1}$ at 430 K for Cu$_{1.8}$Se. The resultant two Lorentzian functions correspond to two diffusion modes with different properties. The FWHM of the narrower Lorentzian function is $Q$ dependent and corresponds to a long-range diffusion mode. The broader Lorentzian function is independent of $Q$ with the FWHM around $7\pm2$ meV. It represents a confined mode that corresponds to short range diffusion between the nearest 8$c$ and 32$f$ sites of Cu ions\cite{NazrulIslam2021}. Adopting one Lorentzian function and a flat background as in previous reports is also tested in our analyses\cite{Danilkin2011,Danilkin2012}. In this way, the FWHMs of the single Lorentzian function are similar to the narrower Lorenzian function corresponding to the long-range diffusion mode. However, the information of the confined mode is lost. The QENS spectra and analyses for the samples at 550 and 650 K are not presented in Fig. \ref{fig2}.

Figure \ref{fig3}(a) displays the FWHMs of the narrower Lorentzian function as a function of $Q$, yielding that the QENS peaks become broader as the temperature increases for Cu$_{1.8}$Se. Increasing of the Cu content also broadens the FWHMs as shown in Figs. \ref{fig3}(b) and \ref{fig3}(c) for Cu$_{1.96}$Se and Cu$_{2}$Se. While the FWHMs for the two compounds with more Cu content are almost temperature independent. An isotropic C-E model is employed to analyze the diffusion process of $\alpha$-Cu$_{2-\delta}$Se\cite{Chudley1961}. In this model, isotropic long-range diffusions can be regarded as a successive jump of mobile atoms between the adjacent crystallographic sites. The C-E model is given by
\begin{equation}
\mathrm{FWHM}(Q)=\frac{2\hbar}{\tau}(1-\frac{\sin(Ql)}{Ql})
\label{eq2}
\end{equation}
where $l$ is the average jump length and $\tau$ is the residence time of an atom at each site. Accordingly, the diffusion coefficient $D$ is expressed as $D=\frac{l^2}{2d\tau}$, where $d=3$ is the dimensional number for the three dimensional diffusion in $\alpha$-Cu$_{2-\delta}$Se. The fits of FWHM to the C-E model are displayed in Figs. \ref{fig3}(a)-\ref{fig3}(c). The resultant temperature dependence of $\tau$, $l$, and $D$ are shown in Figs. \ref{fig3}(d)-\ref{fig3}(f).

The residence time $\tau$ in Fig. \ref{fig3}(d) is on the scale of $10^{-12}$ s in agreement with previous reports\cite{Danilkin2011,Danilkin2012,Voneshen2017,NazrulIslam2021,Kumar2022}. As temperature increases, $\tau$ is expected to become shorter due to the thermal fluctuations as it is indeed observed in Cu$_{1.8}$Se. However, the $\tau$ for Cu$_{1.96}$Se and Cu$_{2}$Se does not change obviously in this temperature range. The temperature dependence of the residence time is related to the diffusion activation energy and may also have connections to the occupancies of the $8c$ and $32f$ sites\cite{NazrulIslam2021}. As one can deduce from the residence time ($\sim$2 ps), the Cu diffusion process is fast and can only affect the phonon spectra below 2 meV. Thus, the phonon anharmonicity or the dynamic creation and destruction of the Frenkel defects may be responsible for the ultra-low thermal conductivity in $\alpha$-Cu$_2$Se\cite{Voneshen2017,NazrulIslam2021}.

The average jump lengths $l$ obtained from the fitting are in the range of $2.7-3.4$ {\AA} as shown in Fig. \ref{fig3}(e). The analysis of the diffusion paths depends on the structural model. In our NPD refinement, the Cu ions are determined to occupy the $8c$ and $32f$ sites. The nearest and next nearest distances for the $8c$ sites of Cu$_{1.96}$Se at 430 K are 2.919(1) and 4.128(1) {\AA}, respectively. These distances are close in the compounds of $\alpha$-Cu$_{2-\delta}$Se ($\delta=0, 0.04, 0.2$). Thus, the hopping between the adjacent $8c$ sites may be the main long-range diffusion mode which can be captured well by the average jump length $l$. Because the occupancy of the 32$f$ site increases as the diffusion coefficient, the 32$f$ site is likely a bypass in the long-range diffusion process\cite{Danilkin2011,Danilkin2012}. The confined mode can be attributed to the localized hopping between the nearest $8c$ and $32f$ sites for Cu ions. Thus, the two diffusion modes cooperate as theoretical simulations\cite{Zhuo2020}.

The diffusion coefficient reflects diffusion rate in a diffusion process. Figure \ref{fig3}(f) displays the diffusion coefficient $D$ as a function of temperature. The magnitude of $D$ on the scale of $10^{-5}$ cm$^2$/s is consistent with most of the QENS results\cite{Danilkin2011,Danilkin2012,NazrulIslam2021,Kumar2022}, while in contrast to that determined from the chemical measurements which are in the magnitudes of $10^{-4}$ or $10^{-6}$ cm$^2$/s\cite{Konev1985,Yakshibaev1984,Korzhuev1989}. Since the average jump length $l$ is nearly temperature independent, the $D$ is enhanced by thermal fluctuations as expected in Cu$_{1.8}$Se. For Cu$_{1.96}$Se and Cu$_{2}$Se, the $D$ is larger, which may result from the stronger correlations between Cu ions in the compounds with higher Cu content. However, the diffusion coefficients do not increase obviously for the two compounds with increasing temperature. The copper occupancy of the 32$f$ sites increases with temperature with a simultaneous decrease in the 8$c$ sites\cite{Skomorokhov2006}. The effect of increasing temperature on the Cu occupancies of the two sites is similar with increasing the nominal Cu content. The enhanced Cu occupancy on the 32$f$ sites by the nominal Cu content may suppress further improving of the Cu occupancy on the same sites by temperature in this range. We note the compositions summarized in Fig. \ref{fig3}(f) are nominal, which may deviate from the actual compositions. Voneshen et. al. separated the diffusion modes in a different way. The diffusion coefficient could not compare with the others directly\cite{Voneshen2017}.

In summary, the average crystal structure and the diffusion behaviors of superionic $\alpha$-Cu$_{2-\delta}$Se ($\delta=0, 0.04, 0.2$) have been studied by NPD and QENS. The Cu ions for $\alpha$-Cu$_{2-\delta}$Se occupy on the tetrahedral $8c$ and trigonal $32f$ sites. A long-range diffusion mode and a confined mode are extracted through the analyses of the QNES spectra. In the long-range diffusion mode, the residence time $\tau$ is on the time scale of picosecond. The jump length $l$ can capture the main jump pathway of hopping between nearest 8c sites for the long-range diffusion. The confined mode corresponds to the localized hoping between the nearest $8c$ and $32f$ sites. The long-range diffusion coefficient is on the scale of $\sim10^{-5}$ cm$^2$/s and increases with elevating temperature for the compounds with a lower Cu content. The diffusion coefficient increases for the compounds with higher Cu content. However, the diffusion rates for compounds with higher Cu content are almost temperature independent, which could be attributed to the compensation of temperature and nominal Cu content on the occupancies of the 8$c$ and 32$f$ sites. The Cu occupancy on the $32f$ site has an obvious connection to the diffusion properties of the long-range mode, consistent with the $32f$ site as a bypass site in the long-range diffusion process. Thus, the two diffusion modes cooperate. These outcomes provide important insights for the understanding of the Cu ions diffusion in superionic $\alpha$-Cu$_{2-\delta}$Se, and shed a light on further investigation  of liquid-like thermoelectric materials.

We thank ANSTO for providing access to Echidna and Pelican (Proposal ID P6942) at the Australian Center for Neutron Scattering. Work at Sun Yat-Sen University was supported by the National Natural Science Foundation of China (Grants No. 12174454, No. 11904414), the Guangdong Basic and Applied Basic Research Foundation (No. 2021B1515120015), and the National Key Research and Development Program of China (Grant No. 2019YFA0705702).
\bibliography{Cu2-xSe}

\end{document}